\documentclass[11pt,a4paper]{article}
\usepackage[utf8]{inputenc}

\usepackage{authblk}
\usepackage{setspace}
\usepackage{complexity}
\usepackage{graphicx}
\usepackage{amsmath}
\usepackage{dsfont}
\usepackage{amsfonts}

\usepackage{subfloat}
\usepackage[subrefformat=parens,labelformat=parens]{subfig}
\usepackage[noend]{algpseudocode}
\algnewcommand{\endIf}{\State\algorithmicend\ \algorithmicif}

\algnewcommand{\endFor}{\State\algorithmicend\ \algorithmicfor}

\algnewcommand{\endFunction}{\State\algorithmicend\ \algorithmicfunction}
\algnewcommand{\endWhile}{\State\algorithmicend\ \algorithmicwhile}

\algnewcommand\algorithmicinput{\textbf{\hskip0.5em input:}}
\algnewcommand\Input{\item[\algorithmicinput]}
\algnewcommand\algorithmicglobal{\textbf{\hskip0.5em global:}}
\algnewcommand\Global{\item[\algorithmicglobal]}
\algnewcommand\algorithmicoutput{\textbf{\hskip0.5em output:}}
\algnewcommand\Output{\item[\algorithmicoutput]}
\algrenewcommand{\algorithmiccomment}[1]{\hskip3em $/^*$\textit{#1}$^*/$}
\makeatletter
\algnewcommand{\lComment}[1]{\Statex \hskip\ALG@thistlm $/^*$\textit{#1}$^*/$}
\makeatother

\algnewcommand{\algorithmicgoto}{\textbf{go to}}%
\algnewcommand{\Goto}[1]{\algorithmicgoto~\ref{#1}}%

\newcommand{\Fbest}{\mathbb{F}_{\text{best}}}
\newcommand{\Fcurr}{\mathbb{F}_{\text{current}}}

\def\abs#1{\mathopen| #1 \mathclose|}
\let\e\varepsilon               

\date{}

\title{Avoiding the Flip Ambiguities in 2D Wireless Sensor Localization by Using Unit Disk Graph Property}
\author{Onur Çağırıcı}

\affil{Technical Report}

\usepackage{parskip}

\usepackage{geometry}
\usepackage{blindtext}

\begin{document}

\maketitle
\begin{abstract}

In this paper, we propose a range-based localization algorithm to localize a wireless sensor network (WSN) in 2D.
A widely used algorithm to localize a WSN in 2D is trilateration, which runs in polynomial time.
Trilateration uses three distance measurements to localize a node.
In some cases, the lack of connectivity leads to a low percentage of localized nodes since a node's position can be fixed using three distance measurements.
We propose an algorithm that finds the position of a node by using the absence of a distance measurement in addition to a third distance measurement.
If two nodes are not able to sense each other, that means the distance between them is more than the sensing range.
Therefore, our algorithm checks if the possible positions of an unlocalized node $u$ is inside the sensing range of a localized node $\ell$ that is not the neighbor of $u$.
In such case, we eliminate one of the possible positions.

\end{abstract}

\section{Introduction}
One of the main challenges in WSN applications is the \textit{localization problem} \cite{adhoc,changthesis,dwrl,survey,gpsfree,gpslesslowcost}.
\textit{Localization} or \textit{positioning} of a WSN is determining the positions of the sensor nodes with respect to each other.
Even though GPS is a very powerful tool to determine the position of an object, it is usually not efficient enough when the sensor nodes are too close to each other.
Considering the cost and the efficiency of the system, \textit{range-based localization} is widely used to determine the positions of the sensor nodes.
Range-based localization is determining the relative positions of the node by only using the measured pairwise distances \cite{laman}.
The localization process is carried out by using a weighted and undirected graph $G=<V,E,W>$, called a \textit{WSN graph}, where each sensor node correspond to a vertex $v \in V$, the pairwise connection between two nodes $v$ and $w$ correspond to an edge $\{v,w\} \in E$ and the computed distance between these nodes correspond to the weight $w(e); e \in E$ of the edge in the graph.
A WSN graph satisfies the property of unit disk graph (UDG) .
In a UDG, an edge $\{v,w\} \in E$ exists if and only if the Euclidean distance between $v$ and $w$ is less than or equal to a specific value called \textit{sensing range}.
Saxe \cite{saxe} showed that localization of a graph in $\mathds{R}^d$ is \NP-complete where $d \in \mathds{Z}^+$, which is closely related to the embeddability problem.
Aspnes \textit{et. al.} than proved that the problem is \NP-complete in \textit{unit disk graphs} as well \cite{unitdisknphard}.

In order to find the position of a point in $\mathds{R}^2$, we need to know the distance of that point to three non-collinear points.
Within the scope of WSN localization, the points are the sensor nodes (or the vertices of the WSN graph), and the distances are the measured distances between sensor nodes (or the weighted edges of the WSN graph).
If there exists an ordering in a WSN graph such that beginning from three points, we are able to determine the positions of the rest, then we can localize all the nodes in the graph in polynomial time, using an algorithm called \textit{trilateration} \cite{rigidity}.
The ordering, hence, is called \textit{trilaterative ordering} and WSN graph that contains such ordering is called \textit{trilateration graph} \cite{theory}.

\begin{figure}[htbp]
\centering
	\subfloat[A non-violating point formation]{%
	\includegraphics[width=0.4\linewidth]{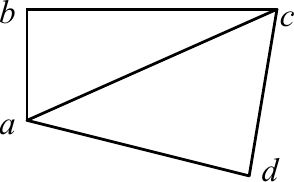}
	\label{fig:forigid}%
	}
	\hfill
	\subfloat[A UDG violation]{%
	\includegraphics[width=0.4\linewidth]{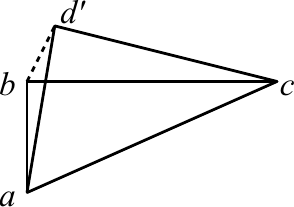}
	\label{fig:forigidw}%
	}
	
\caption{When we fix the positions of $a$, $b$ and $c$, (a) does not lead to a violation but (b) is a violation}
\label{fig:missingEdge}
\end{figure}

We exploit UDG property to eliminate one of the possible positions for an unlocalized node that has two localized neighbors.
Instead of looking for a third distance measurement to localize a node, we forbid a node to be in the area that is covered by the sensing ranges of the localized nodes that are not neighbors of the unlocalized node.
We show such a case in Figure \ref{fig:missingEdge}.

In Figure~\subref*{fig:forigid}, we see four nodes with five pairwise distances.
By definition, the graph is not globally rigid and not localizable.
However, by a simple reasoning, we can observe that the point formation shown in Figure~\subref*{fig:forigidw} is violating the UDG property because there is no such edge $\{b,d'\}$.

\section{Assumptions and Terminology}
We have mentioned the concepts of WSN graph and UDG in the previous section.
Before moving onto the algorithm, let us introduce the assumptions and terminology that we use throughout the manuscript in this section.

We assume that all the sensors are fully functional and identical i.e. the graphs that we deal with are UDGs.
This assumption forms the base of our algorithm.
The neighbor list $N(i)$ of a node $i$ contains another node $j$ if and only if the Euclidean distance between $i$ and $j$ is less than or equal to the sensing range of the sensors $\Theta$.
Neighbor list of a node is an ordered list and $j^{th}$ neighbor of the node $i$ is denoted by $N_j(i)$.
The last neighbor in the neighbor list of $i$ is denoted by $N_\textit{last}(i)$.
Similarly, a node $i$ has a list of its localized neighbors denoted by $LN(i)$.
Following the same notation, $LN_j(i)$ is the $j^{th}$ localized neighbor of $i$ and $LN_\textit{last}(i)$ denotes the most recently localized neighbor of $i$.
The number of localized neighbors of $i$ is also kept and denoted by $i.\textit{LNCount}$.

The edge weights of the network graph $G$ is determined by adding a random value $\e \in \R$ to each actual pairwise distance $\delta^\text{actual}_{ij}$.
Thus, the pairwise distances are computed as $\delta_{ij} = \delta^\text{actual}_{ij} + \e$, where $\e$ is the randomly gathered noise value unless stated otherwise.
The error model will be detailed in Section~\ref{sec:experiments}.
In order to denote the measured distance between a node $i$ and its $j^{th}$ neighbor, instead of $\delta_{iN_j(i)}$, we use the notation $\delta_{N_j(i)}$.

A \textit{unit disk graph violation} is placing a node $i$ at some position $(x,y)$ such that the Euclidean distance between the position of a localized node $\ell \not\in N(i)$ and $(x,y)$ is smaller than or equal to the sensing range.

\section{Localization a WSN graph in 2D Using UDG Violations}

When a node is being localized, two distance measurements leave us with two possible positions for that node.
Hence, we need a third distance measurement in order to fix the position of the node.
Considering the noise, we give the function to choose between two possible points in Figure~\ref{func:choose}.

\begin{figure}[htbp]
\begin{align*}
	\sigma(p_1, p_2, d) = \left\{
	\begin{array}{ll}
	\textsc{null}, &\text{if } (\abs{\delta_{jp_{1}} - d} \leq d*P) \bigwedge
	(\abs{\delta_{jp_{2}} - d} \leq d*P)\\
	p_1, &\text{if }\abs{\delta_{jp_{1}} - d} \leq d*P\\
	p_2, &\text{if } \abs{\delta_{jp_{2}} - d} \leq d*P\\
	\textsc{null}, &\text{otherwise} 
	\end{array}
	\right.
	\end{align*}
	\caption{Picking a position among two candidate points}
	\label{func:choose}
\end{figure}

\begin{figure}[htbp]
\fbox{
\begin{minipage}{0.9\columnwidth}
\textsc{LocalizeGraph($G$)}
\begin{algorithmic}[1]
\Input{WSN graph $G = <V,E,W>$}
\Output{2D point formation of $G$}
\State $\Fbest \gets ([\text{  }],[\text{  }])$ \Comment{Best point formation} \label{initFbest}
\For{each non-collinear and fully-connected $\{a,b,c\} \subseteq V$} \label{trilatIterateSeed}
\State $\Fcurr \gets  ([\text{  }],[\text{  }])$ \label{initFcurr}
	\State Pick $a$, $b$ and $c$ as seed nodes
	\State $Q_{\textit{process}} \gets [a,b,c]$ \label{initQ}
	\While{$Q_{\textit{process}}$ is not empty} \label{while}
		\State $i \gets \textit{dequeue}(Q_{\textit{process}})$ \label{dequeue}
		\State add $\left(i, i.\textit{Pos}\right)$ into $\Fcurr$ \label{addIntoF}
		\For{all $j \in N(i)$} \label{iterateOnNeighbors}
			\State $\textit{result} \gets \textsc{false}$
			\If{$j.\textit{LNCount} \geq 2$}
				\State $\textit{result} \gets$ \Call{Bilaterate}{$j$}
			\EndIf
			\If{$j.\textit{LNCount} \geq 3$}
				\State $\textit{result} \gets$ \Call{Trilaterate}{$j$}
			\EndIf
			\If{$\textit{result} = \textsc{true}$}
			\State $\textit{enqueue}(j,Q_{\textit{localized}})$ \label{trilatAddIntoQ}
			\EndIf
		\EndFor
	\EndWhile
	\endWhile \label{trilatEndWhile}
	\If{$|\Fcurr^1| = |V|$} 
		\Return $\Fcurr$ \label{trilatAllLocalized}
	\EndIf
	\If{$|\Fcurr^1| > |\Fbest^1|$}
		$\Fbest \gets \Fcurr$ \label{trilatFisBetter}
	\EndIf
\EndFor
\endFor \label{trilatIterateSeedEnd}
\State \Return $\mathbb{F}_\text{best}$ \label{trilatReturnBest}
\end{algorithmic} \end{minipage}
}
\caption{Overall localization algorithm}
\label{alg:trilatquad}
\end{figure}

\begin{figure}[htbp]
\fbox{
\begin{minipage}{\columnwidth}
\textsc{Bilaterate($j$)}
\begin{algorithmic}[1]
\State Find possible positions $p_1, p_2$ of $j$ with respect to $LN(j)_1$, $LN_\textit{last}(j)$
\State $\textit{violate}_1 \gets \textsc{false}$
\State $\textit{violate}_2 \gets \textsc{false}$
\For{each localized node $i \not\in N(j)$}
	\If{$d(p_1,i) < \Theta$}
		$\textit{violate}_1 \gets \textsc{true}$
	\EndIf
	\If{$d(p_2,i) < \Theta$}
		$\textit{violate}_2 \gets \textsc{true}$
	\EndIf
	\If{$(\textit{violate}_1)$ AND $(\textit{violate}_2)$}
		\Return \textsc{false}
	\EndIf
\EndFor
\If{$(!\textit{violate}_1)$ AND $(!\textit{violate}_2)$}
	\Return \textsc{false}
\EndIf
\If{$!\textit{violate}_1$}
	$j.\textit{Pos} \gets p_1$
\EndIf
\If{$!\textit{violate}_2$}
	$j.\textit{Pos} \gets p_2$
\EndIf
\State \Return \textsc{true}
\end{algorithmic} \end{minipage}
}
\caption{Bilateration with missing edges}
\label{alg:bilat}
\end{figure}

\begin{figure}[htbp]
\fbox{
\begin{minipage}{\columnwidth}
\textsc{Trilaterate($j$)}
\begin{algorithmic}[1]
\State Find possible positions $p_1, p_2$ of $j$ with respect to $LN_1(j)$, $LN_2(j)$ and $LN_\textit{last}(j)$

\State $\textit{violate}_1 \gets \textsc{true}$
\State $\textit{violate}_2 \gets \textsc{true}$
\For{each localized node $i \not\in N(j)$}
	\If{$d(p_1,i) < \Theta$}
		$\textit{violate}_1 \gets \textsc{false}$
	\EndIf
	\If{$d(p_2,i) < \Theta$}
		$\textit{violate}_2 \gets \textsc{false}$
	\EndIf
	\If{$(!\textit{violate}_1)$ AND $(!\textit{violate}_2)$}
		\Return \textsc{false}
	\EndIf
\EndFor
\endFor
\If{$(!\textit{violate}_1)$ AND $(!\textit{violate}_2)$}
	\State $j.\textit{Pos} \gets $ \Call{$\sigma$}{$p_1, p_2, LN_\textit{last}(j)$}
	\If{$j.Pos$ is \textsc{null}}
		\Return \textsc{false}
	\EndIf
	\State \Return \textsc{true}

\EndIf
\endIf

\If{$!\textit{violate}_1$}
	$j.\textit{Pos} \gets p_1$
\EndIf
\If{$!\textit{violate}_2$}
	$j.\textit{Pos} \gets p_2$
\EndIf

\State \Return \textsc{true}
\end{algorithmic} \end{minipage}
}
\caption{Trilateration with missing edges}
\label{alg:trilat}
\end{figure}

\newpage
\section{Localizing a Wheel Graph in Polynomial Time}
In this section, we explain how our algorithm can be used to localize a wheel graph in polynomial time.
Clearly, we assume that the corresponding graph is a unit disk graph.
Aspnes \textit{et al.} proved that localizing a wheel graph is \NP-hard \cite{theory}.
The proof is reduction from SET COVER problem.
Given the fact that a WSN graph should satisfy the property of a UDG, we then are able to use our algorithm to localize a wheel graph.
If we pick three nodes as seed, then a fourth node can be localized by using two of the seed nodes.
Aspnes \textit{et al.} \cite{theory} mentions that since we are not able to know which direction to place the nodes of a wheel graph, the problem is hence \NP-hard.
Considering UDG violations, we are able to fix the positions of the nodes in polynomial time with the algorithm given in Figure~\ref{alg:bilat}.

\begin{figure}[htbp]
\centering
\includegraphics[width=0.5\linewidth]{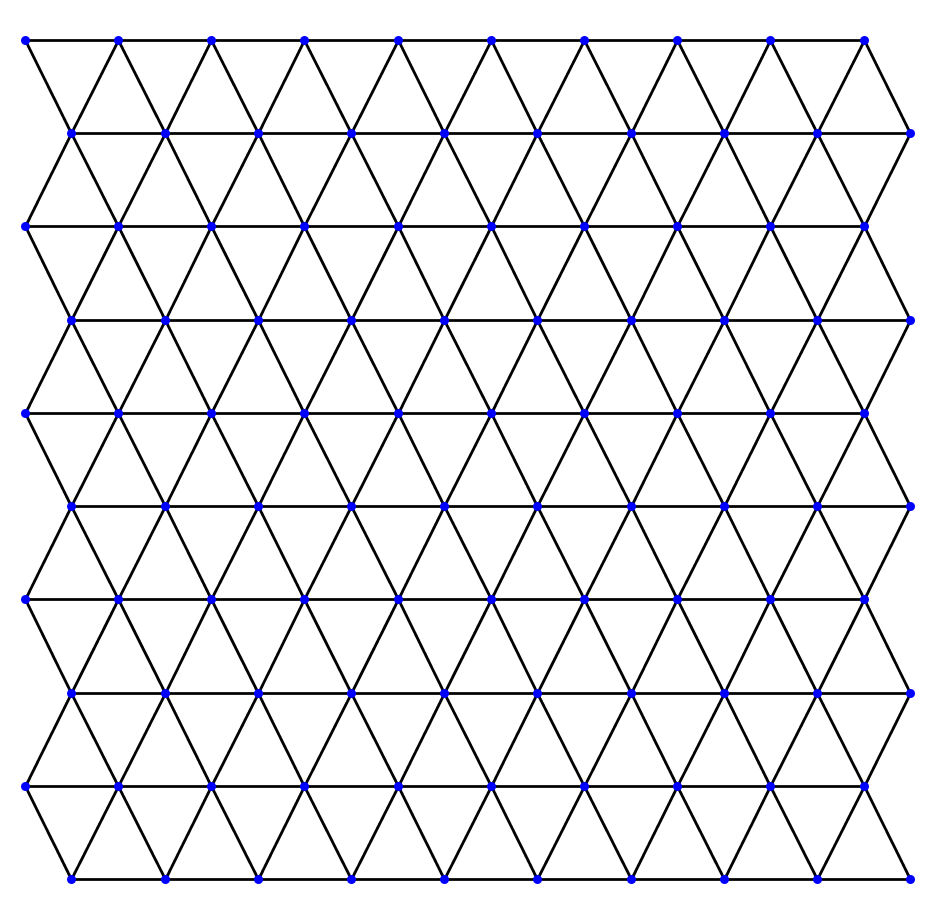}
\caption{A wheel graph}
\label{fig:wheel}
\end{figure}

\begin{figure}[htbp]
	\centering
	\subfloat[$P=1$]{%
		\includegraphics[width=0.45\linewidth]{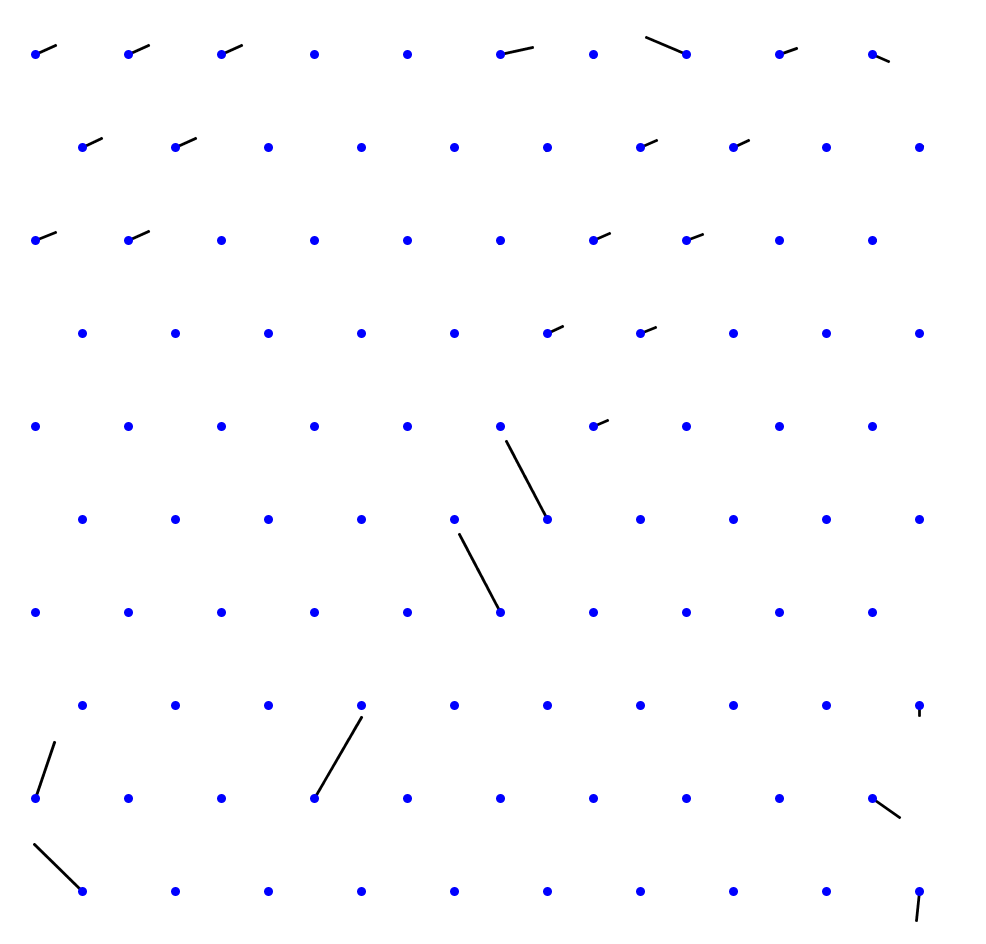}
		\label{fig:wheel1}%
	}
	\hfill
	\subfloat[$P=5$]{%
		\includegraphics[width=0.45\linewidth]{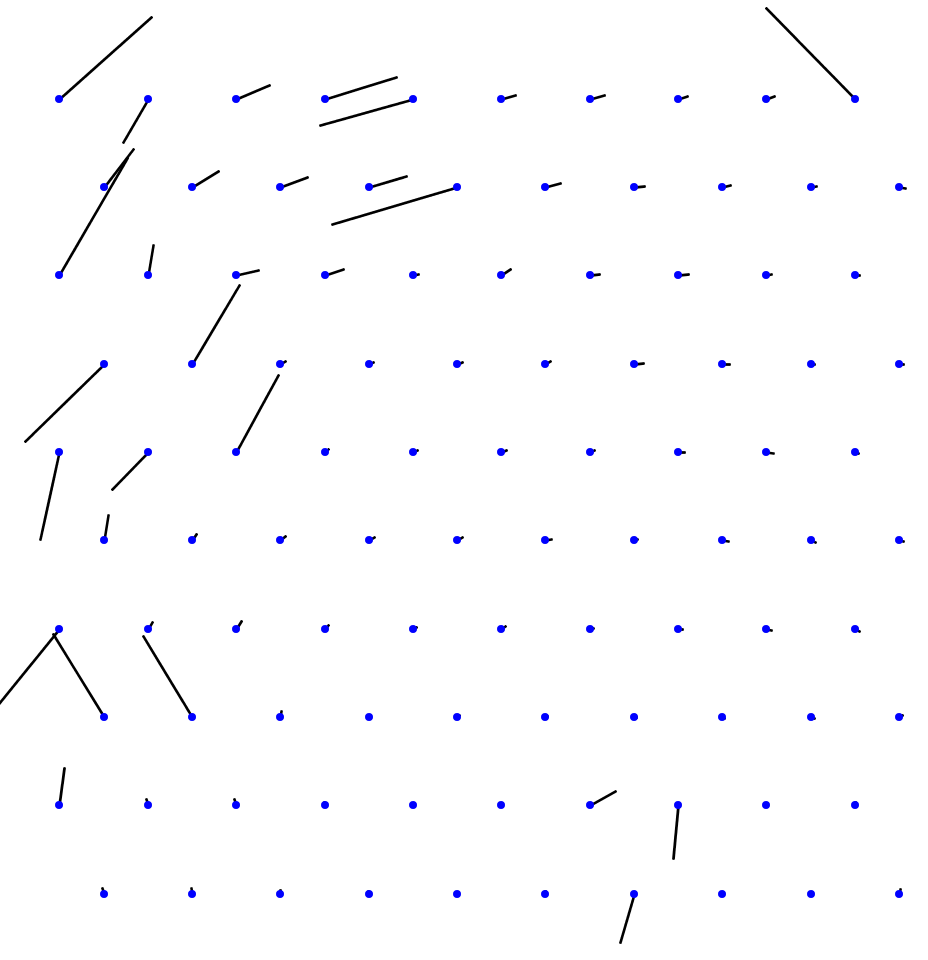}
		\label{fig:wheel5}%
	}

	\subfloat[$P=10$]{%
		\includegraphics[width=0.45\linewidth]{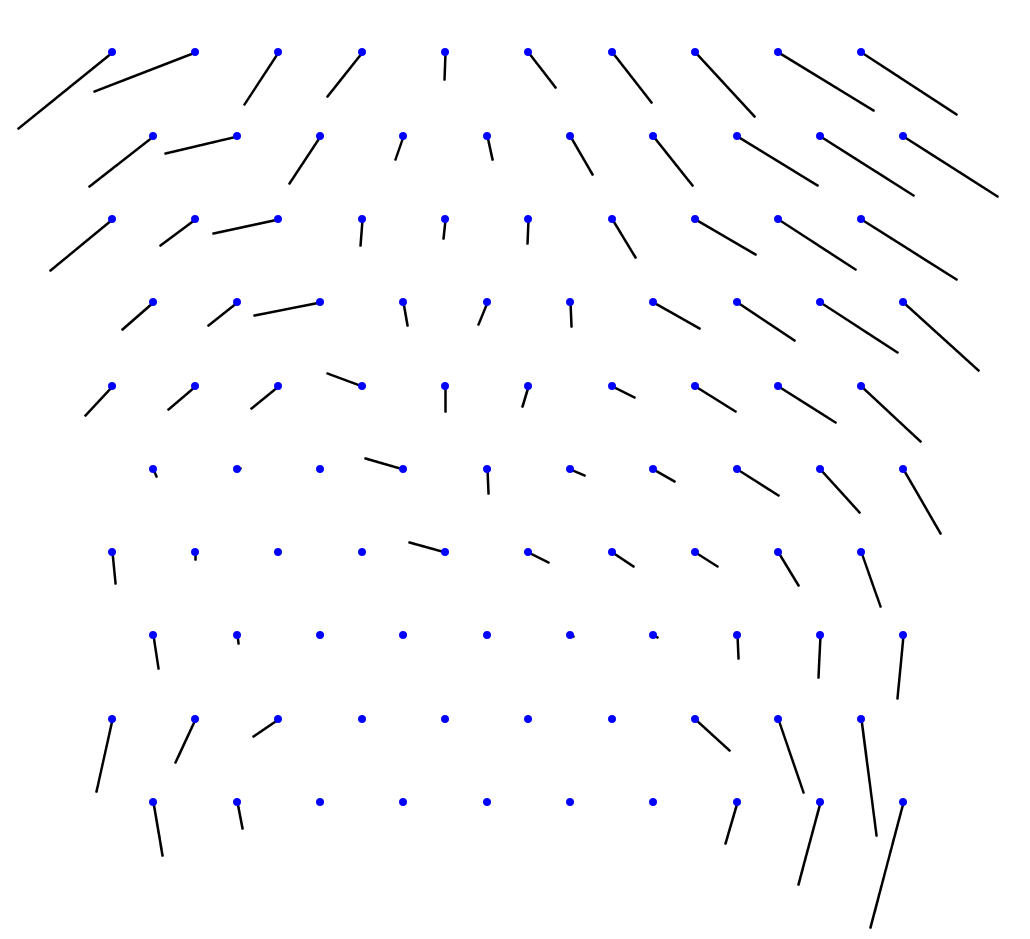}
		\label{fig:wheel10}
	}
	\hfill
	\subfloat[$P=20$]{%
		\includegraphics[width=0.45\linewidth]{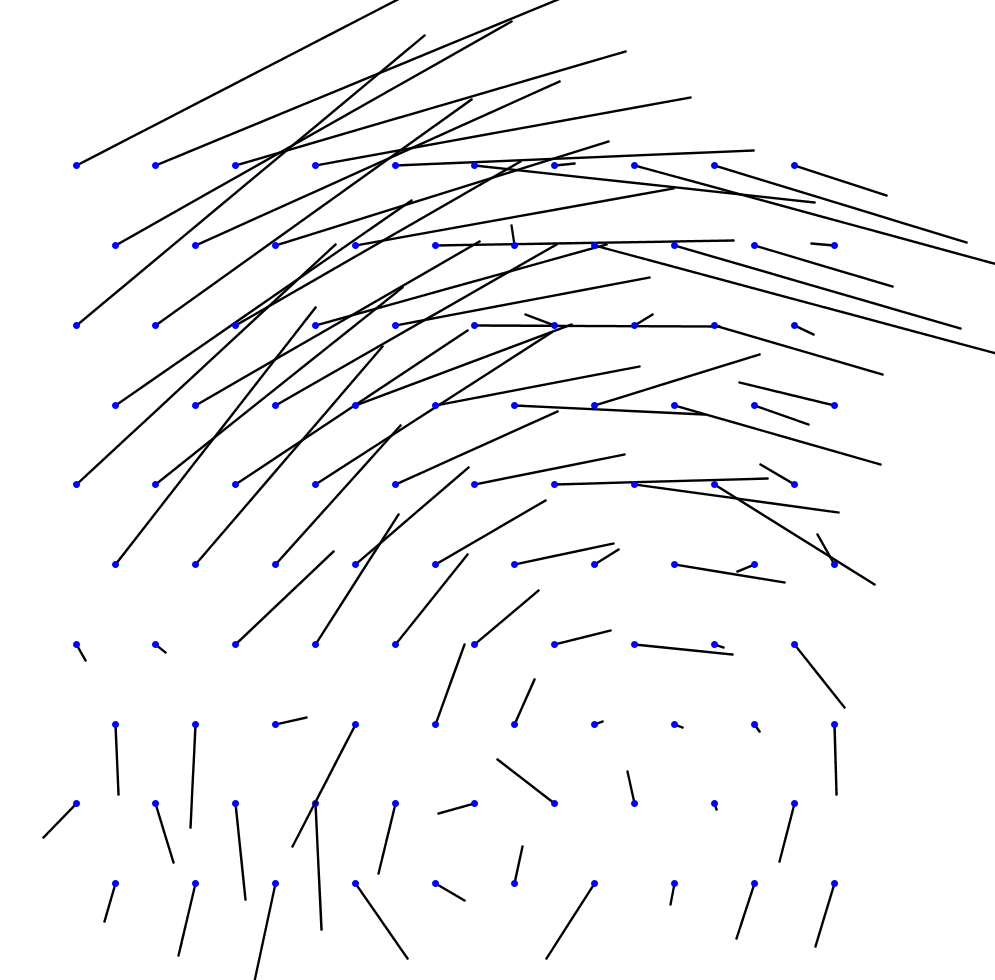}
		\label{fig:wheel20}%
	}
	
	\caption{Localization of a WSN that contains many wheel graphs where error magnitude is 1 (a), 5 (b), 10 (c) and 20 (d)}
	\label{fig:wheelGraphLocalizatiohn}
\end{figure}

In Figure~\ref{fig:wheel} we see a WSN graph that is localizable but does not have a trilaterative ordering.
Since the graph contains many wheel graphs, trilateration cannot localize any of the nodes in this graph.
In Figure~\ref{fig:wheelGraphLocalizatiohn}, we give four examples of localization of the graph given in Figure~\ref{fig:wheel} with various values of noise magnitude.

\newpage
\section{Experimental Results} \label{sec:experiments}
In this section, we conduct experiments to see the how our algorithm improves localization with respect to the mere trilateration.
We compare two algorithms first with noiseless distance measurements to see the effect of connectivity on both.
Then, we pick the connectivity values where each algorithm reached to a recall percentage greater than $99\%$ and see the effect of environmental noise.
The noise is modeled inline with experimentally gathered data in \cite{realworld1, realworld2}.
Each edge $\{v,w\} \in E$ in $G=(V,E)$ is modified with respect to a random number generated using a Gaussian random distribution with $N(f(\Theta), P/100)$ where $P$ is the magnitude of the error and $\Theta$ is the sensing range.
$f(\Theta)$ is defined as follows;

\begin{align*}
f(\Theta) = 0.022 ln(1+\Theta) - 0.038
\end{align*}

\begin{figure}
\centering
\includegraphics[width=0.7\linewidth]{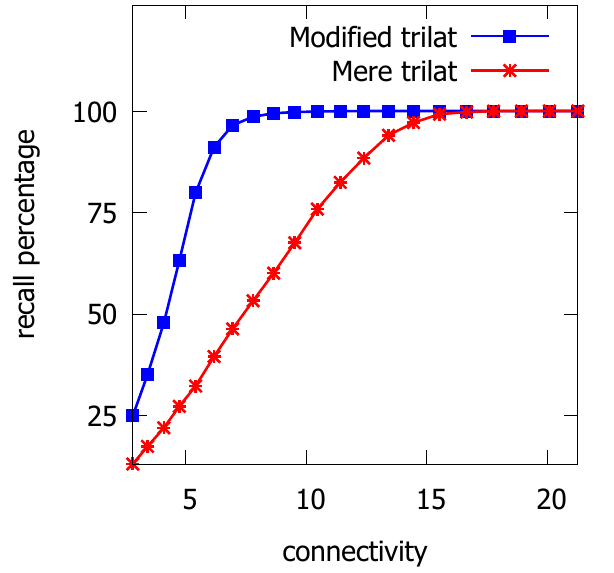}
\caption{Recall percentage of trilateration when UDG violations are considered and when they are not.}
\label{fig:recall}
\end{figure}

In Figure~\ref{fig:recall} we show the results of tests with noiseless range measurements.
The graph tells us that if UDG violations are considered, a WSN graph can be localized with less sensing range.
Our algorithm is able to localize more than $99\%$ of the nodes when the connectivity is around 8 whereas mere trilateration requires 15 connectivity per node.

\begin{figure}
	\centering
	\includegraphics[width=0.7\linewidth]{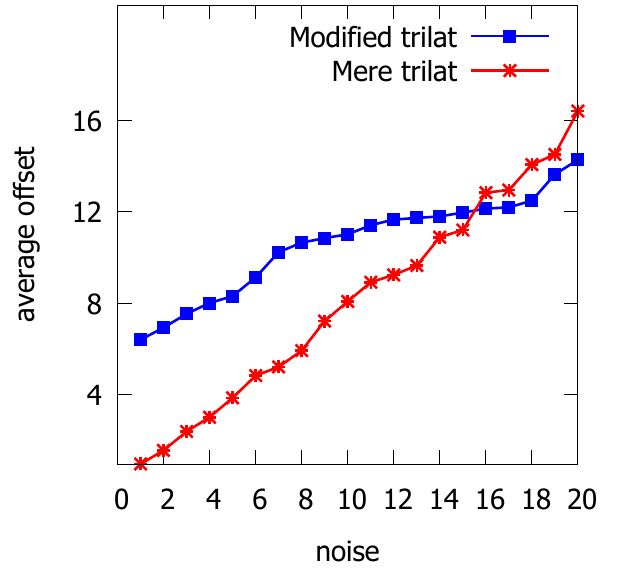}
	\caption{Average offset of trilateration when UDG violations are considered and when they are not.}
	\label{fig:offset}
\end{figure}

Now, let us show the effect of environmental noise.
In Figure~\ref{fig:offset}, we set the connectivity value as 8 units for our algorithm and 15 units for mere trilateration and plot the average offsets of both algorithms.
The graph tells us that considering UDG violations causes more offset than mere trilateration.

\newpage
\begin{spacing}{0.9}
	\bibliographystyle{ieeetr}
	\bibliography{ref}

\begin{thebibliography}{10}

\bibitem{adhoc}
S.~Capkun, M.~Hamdi, and J.~Hubaux, ``{GPS-free positioning in mobile ad-hoc
  networks},'' in {\em System Sciences, 2001. Proceedings of the 34th Annual
  Hawaii International Conference on}, p.~10, 2001.

\bibitem{changthesis}
C.~Chang, ``{Localization and Object-Tracking in an Ultrawideband Sensor
  Network},'' Master's thesis, UC Berkeley, USA, 2004.

\bibitem{dwrl}
H.~Akcan and C.~Evrendilek, ``{GPS-free directional localization via dual
  wireless radios},'' {\em Computer Communications}, vol.~35, no.~9,
  pp.~1151--1163, 2012.

\bibitem{survey}
A.~Pal, ``{Localization Algorithms in Wireless Sensor Networks: Current
  Approaches and Future Challenges},'' {\em Network Protocols and Algorithms},
  vol.~2, no.~1, pp.~45--73, 2010.

\bibitem{gpsfree}
H.~Akcan, V.~Kriakov, H.~Bronnimann, and A.~Delis, ``{GPS-Free node
  localization in mobile wireless sensor networks},'' in {\em Proceedings of
  the 5th ACM International Workshop on Data Engineering for Wireless and
  Mobile Access (MobiDE'06)}, (Chicago, Illinois, USA), p.~35–42, 2006.

\bibitem{gpslesslowcost}
N.~Bulusu, J.~Heidemann, and D.~Estrin, ``{GPS-less low-cost outdoor
  localization for very small devices},'' {\em Personal Communications, IEEE},
  vol.~7, pp.~28--34, Oct 2000.

\bibitem{laman}
G.~Laman, ``{On graphs and rigidity of plane skeletal structures},'' {\em
  Journal of Engineering Mathematics}, vol.~4, no.~10, pp.~331 -- 340, 2002.

\bibitem{saxe}
J.~Saxe, ``{Embeddability of weighted graphs in k-space is strongly
  {NP}-Hard.},'' in {\em 17th Allerton Conference in Communications, Control
  and Computing on}, pp.~480--489, 1979.

\bibitem{unitdisknphard}
J.~Aspnes, D.~Goldenberg, and R.~Yang, ``{On the Computational Complexity of
  Sensor Network Localization},'' in {\em In Proceedings of First International
  Workshop on Algorithmic Aspects of Wireless Sensor Networks}, pp.~32--44,
  2004.

\bibitem{rigidity}
T.~Eren, O.~Goldenberg, W.~Whiteley, Y.~R. Yang, A.~Morse, B.~D.~O. Anderson,
  and P.~Belhumeur, ``{Rigidity, computation, and randomization in network
  localization},'' in {\em INFOCOM 2004. Twenty-third AnnualJoint Conference of
  the IEEE Computer and Communications Societies}, vol.~4, pp.~2673--2684,
  2004.

\bibitem{theory}
J.~Aspnes, T.~Eren, D.~Goldenberg, A.~Morse, W.~Whiteley, Y.~Yang, B.~Anderson,
  and P.~Belhumeur, ``{A theory of network localization},'' in {\em {IEEE}
  Transactions on Mobile Computing}, vol.~5, pp.~1663--1678, 2010.

\bibitem{realworld1}
B.~Alavi and K.~Pahlavan, ``{Modeling of the TOA-based distance measurement
  error using UWB indoor radio measurements},'' {\em Communications Letters,
  IEEE}, vol.~10, pp.~275--277, Apr 2006.

\bibitem{realworld2}
J.~Park, E.~Demaine, and S.~Teller, ``{Moving-Baseline Localization},'' in {\em
  Information Processing in Sensor Networks, 2008. IPSN '08. International
  Conference on}, pp.~15--26, Apr 2008.

\end{thebibliography}
\end{spacing}
\end{document}